\documentclass[structabstract]{aa}
\usepackage{graphicx}
\usepackage{natbib}
\usepackage{txfonts}
\begin{document}
   \title{An iterative method in a probabilistic approach to the spectral inverse problem}

   \subtitle{Differential emission measure from line spectra and broadband data}

   \author{F.~F. Goryaev\inst{1,2}, S. Parenti\inst{1}\fnmsep\thanks{Now at the Institut d'Astrophysique Spatiale, Paris-Sud University, 91405 Orsay Cedex, France}, A.~M. Urnov\inst{2,3}, S.~N. Oparin\inst{2}, J.-F. Hochedez
          \inst{1},
          \and
          F. Reale\inst{4}
          }

   \institute{
              Royal Observatory of Belgium, Avenue Circulaire 3, 1180 Brussels, Belgium\\
              \email{farid.goryaev@oma.be}
         \and
             Lebedev Physical Institute, Russian Academy of Sciences, Leninskii pr. 53, Moscow, 119991 Russia\\
                     \and
                     Moscow Institute of Physics and Technology, Dolgoprudnii, Moscow region, 141700 Russia\\
                     \and
                     Dipartimento di Scienze Fisiche \& Astronomiche, Universita di Palermo, Piazza del Parlamento 1, 90134 Palermo, Italy
             }

   \date{Received ... / Accepted ...}

  \abstract
   {Inverse problems are of great importance in astrophysics, e.g., for deriving information about the physical characteristics of hot optically thin plasma sources from their extreme ultraviolet and X-ray spectra.
   }
   {We describe and test an iterative method developed within the framework of a probabilistic approach to the spectral inverse problem for determining the thermal structures of the emitting plasma. We also demonstrate applications of this method to both high resolution line spectra and broadband imaging data.
   }
   {Our so-called Bayesian iterative method (BIM) is an iterative procedure based on Bayes' theorem and is used to reconstruct differential emission measure (DEM) distributions.
   }
   {To demonstrate the abilities of the BIM, we performed various numerical tests and model simulations establishing its robustness and usefulness. We then applied the BIM to observable data for several active regions (AR) previously analyzed with other DEM diagnostic techniques: both SUMER/SOHO (Landi \& Feldman 2008) and SPIRIT/CORONAS-F (Shestov et al. 2010) line spectra data, and XRT/Hinode (Reale et al. 2009) broadband imaging data. The BIM calculations confirmed the main results for SUMER/SOHO data showing very good quantitative agreement between both DEMs at $\log T \approx 6.5$ ($T$ is the temperature in units of Kelvin) and a slight shift for two maxima at lower temperatures with $\approx$ 30--50\% difference in the DEM values for the coolest peak. For the SPIRIT data, we revised and validated AR DEM results including the inference of hot plasma in ARs with an emission measure (EM) at temperatures $\approx$ 9--15 MK comparable to the EM at $\approx$ 2--4 MK. In the case of XRT broadband data, the BIM solutions provided evidence of hot plasma at temperatures $\approx$ 4--6 MK with EM up to $\sim$ 30\% as compared to that at $\approx$ 2--4 MK in a non-flaring AR on 2006 November 12.
   }
   {The BIM results show that this method is an effective tool for determining the thermal structure of emitting plasma and can be successfully used for the DEM analysis of both line spectra and broadband imaging data. The BIM calculations correlate with recent studies confirming the existence of hot plasma in solar ARs. The BIM results also indicate that the coronal plasma may have the continuous distributions predicted by the nanoflare paradigm.
   }


   \titlerunning{An iterative method in a probabilistic approach to the spectral inverse problem}
   \authorrunning{F.~F. Goryaev et al.}
   \maketitle
%

\section{Introduction}
\label{intro}

A knowledge about the plasma temperature content of the solar corona is crucially important to understanding the coronal heating problem, mechanisms of explosive and eruptive phenomena
(flares, CMEs, jets, etc.), and the roles of magnetic reconnection and waves in energy release processes. In particular,
considerable interest in the detection of hot plasma ($T>5$ MK) in coronal loops has arisen (Zhitnik et al. 2006; Urnov et al. 2007b; Reale et al. 2009; Schmelz et al. 2009a, 2009b; Patsourakos \& Klimchuk 2009; Shestov et al. 2010). This finding may be related to a coronal heating mechanism, based on the Parker's idea of nanoflares (Parker 1988). Nanoflares are defined as short impulsive energy releases caused by reconnection events in individual magnetic flux strands (see, e.g., Cargill 1994, Klimchuk 2006). A persistent hot, but faint, plasma component is required in the nanoflare paradigm. Its detection and quantification is important to validate (or to discard) this paradigm. For this reason, the availability of reliable diagnostics methods using the observational data is important.

Extreme ultraviolet (EUV) and soft X-ray (SXR) emission from the Sun is one of the main sources of information about the coronal plasma characteristics, needed to constrain the classes of relevant plasma models and enable quantitative simulations of plasma processes -- space-time dynamics of electron temperature $T$ and density $N_e$ in the emitting plasma regions (see Urnov et al. 2007b). The most general quantity that can be inferred from the total flux intensities in spectral lines or broadband channels, measured by means of optically thin XUV spectra and imaging spectroscopy, is a bivariate differential emission measure (DEM) -- the density distribution function for the emission measure (EM), $\mu(T,N_e)$, over both variables $T$ and $N_e$ (Jefferies et al. 1972; Brown et al. 1991). This distribution function can be used to derive various plasma parameters of interest, such as the multithermal characteristics of the emitting plasma or its mean temperature and density. The reconstruction of the bivariate DEM is a rather problematical task of the spectral inverse problem. However, in some special cases the emission measure differential in temperature, DEM($T$) (i.e., the function $\mu(T,N_e)$ integrated over the density $N_e$), can be obtained. A knowledge of the DEM($T$) distribution and corresponding EM function, characterizing the distribution of ``quantity of matter'' over the temperature, makes it possible to study the temperature content of the emitting plasma structures.

There are many methods for performing the DEM distribution reconstruction from calibrated spectroscopic data (see, e.g., Phillips et al. 2008). The basic (``standard'') approach traditionally used formulates the inverse problem in terms of the Fredholm integral equation relative to the volume or column (in the case of imaging spectroscopy with high spatial resolution) DEM distribution. In the present paper, a probabilistic approach based on another mathematical formalization of the spectral inverse problem is applied (see, e.g., Tarantola 2005). In this approach, relative fluxes and normalized functions are treated as probability distributions of the random variables, and an iterative procedure is followed in a regular way based on Bayes' theorem (Bayesian Iterative Method, BIM). In contrast to the traditionally used formalization of the inverse problem within the framework of the Fredholm integral equation or system of algebraic equations, the problem formalized in the probabilistic approach is not ``ill-posed'', i.e., the regularization procedure is not needed (see Sect. \ref{sect_2}). One of the BIM features is that its algorithm, which is assumed to be a sequential estimate based on a current hypothesis for the DEM distribution, corresponds to the maximum likelihood (ML) criterion providing the ultimate resolution enhancement relative to other non-parametric methods (see, e.g., Gelfgat et al. 1993 and references therein).

We note that there exist ``intermediate'' DEM reconstruction methods formulated within the standard approach, but adopting statistical methods to analyze the observed data.
For instance, Kashyap \& Drake (1998) developed and applied a DEM technique by employing a Markov-chain Monte Carlo (MCMC) algorithm, that is based on a Bayesian formalism. This technique allows its users to estimate the most probable DEM distribution describing the observed line fluxes (see also Sect. \ref{sect_2}).

The BIM has proven to be helpful in solving a series of problems such as the image restoration (Richardson 1972) and the signal recovery of noisy data (Gelfgat et al. 1993). It was successfully applied to the deconvolution of initial X-ray spectra recorded by Bragg spectrometers onboard the ``Vertikal-9'' rocket in 1981 (Zhitnik et al. 1987), and also the DEM temperature analysis of both X-ray spectra from tokamak plasma (Goryaev et al. 2003; Urnov et al. 2007a) and EUV spectra observed in the SPIRIT experiment onboard the CORONAS-F satellite (Zhitnik et al. 2006; Urnov et al. 2007b). The application of the BIM to the spectra obtained onboard the CORONAS-F in Zhitnik et al. (2006) and Urnov et al. (2007b) made it possible for the first time to detect in a number of active regions (AR) hot plasma in the temperature range of $\log T \sim 6.8-7.2$ (see Figs. 6 and 7 in Zhitnik et al. 2006, and Fig. 4 in Urnov et al. 2007b). In Shestov et al. (2010), this result was reproduced on the basis of twelve AR spectra and confirmed by the image of an AR in the Mg XII ion resonance line ($\lambda = 8.42$ \AA ) recorded simultaneously in the same SPIRIT experiment. A few preliminary simple tests demonstrating some abilities of the BIM algorithm were also given in Shestov et al. (2010).

The present paper represents a deeper investigation and more detailed analysis of the BIM abilities for the DEM reconstruction by means of numeric tests and model simulations in the case of both line spectra and broadband imaging data. Applications to both high resolution line spectra (provided by such instruments as SUMER onboard the SOHO and SPIRIT on the CORONAS-F satellites) and  broadband spectroscopic data (such as those provided by the soft X-ray telescope (XRT) onboard the Hinode satellite) are also considered. Flux values in a few wide bands generally provide weaker constraints on DEM curves than the DEM analysis applied to line spectra. Through our tests and model simulations, we show that the BIM can be effectively applied to derive the DEM even for XRT broadband data, if they come from a sufficiently large number of different filters.

In Sect. \ref{sect_2}, we formulate the probabilistic approach to the spectral inverse problem and describe the BIM algorithm used for the DEM analysis. In Sect. \ref{sect_3}, numerical tests and model simulations are presented establishing the BIM's robustness and stability. In Sect. \ref{sect_4}, we apply the BIM to the DEM analysis of spectroscopic data from SUMER (Landi \& Feldman 2008), SPIRIT (Shestov et al. 2010), and XRT (Reale et al. 2009) ARs and compare them with other diagnostic techniques. In Sect. \ref{sect_5}, the results are discussed and some conclusions are given.

\section{Probabilistic approach to the spectral inverse problem}
\label{sect_2}

The total power of emission $F(l,\Delta T)$ $\mathrm{[erg\, s^{-1}\, sr^{-1}]}$ in the wavelength band $\Delta\lambda(l)$ of the spectral channel $l$ produced by the temperature interval $\Delta T$ from an optically thin plasma region with the volume $V$ may be written as (see Urnov et al. 2007b and references therein)

\begin{equation}
       F(l,\Delta T)= \int\limits_{V}  G\left( l,T(\mathbf{r})\right) N_e^2(\mathbf{r})\, d\mathbf{r} \, , \label{flux1}
\end{equation}
where $T(\mathbf{r})= \phi(\mathbf{r})$ is a single-valued temperature function in the volume $V$, $N_e(\mathbf{r})$ is the electron density distribution, and $G(l,T)$ is the temperature response function defined as the spectral luminosity distribution function $\widetilde{F}(\lambda ;T,N_e)$ calculated using definite model assumptions about emitting plasma (including the coronal approximation and the optically thin condition) and integrated over $\Delta\lambda(l)$ with the known filter (``apparatus'') function $f(l,\lambda)$

\begin{equation}
          G(l,T) = {1\over N^2_e} \int\limits_{\Delta\lambda(l)} f(l,\lambda ) \widetilde{F}(\lambda ;T,N_e)\, d\lambda \, .
\end{equation}
For a line spectrum, the function $f(l,\lambda)=1$ and $G(l,T)$ coincides with the so-called contribution function of the line $l$.

For the spectroscopic problem that we consider here, which is based on the mathematical formalization (model) of the physical problem, the measuring quantity $F(l,\Delta T)$ can be formally represented as the Stieltjes integral over temperature $T$ in the interval $\Delta T$ (Urnov et al. 2007b)

\begin{equation}
       F(l,\Delta T)= \int\limits_{\Delta T}  G(l,T)\, dY(T) \, , \label{flux2}
\end{equation}
for the volume emission measure (EM) $Y(T)$, which represents an integrating function (non-decreasing) that is the distribution function of matter over temperature.
If the temperature gradient becomes zero ($\nabla\phi =0$) in regions with volume $V_i\in V$ and temperature $T_i$, then the integral in Eq. (\ref{flux2}) can
be broken into the sum of the integrals over regions $V_i$ and a region with $T\in \Delta T$, where $\phi$ is a piecewise smooth function with the gradient $\nabla\phi\neq 0$ in the volume $\widetilde{V}=V-\sum\limits_i V_i$.
The temperature profile of the volume EM in Eq. (\ref{flux2}) can then be represented as the sum of the two terms

\begin{equation}
       Y(T)= Y_c(T) + Y_s(T) \, ,
\end{equation}
described, respectively, by piecewise smooth, $Y_c(T)$, and discontinuous, $Y_s(T)$, non-decreasing functions,

\begin{equation}
     Y_c(T)=\int\limits_{\widetilde{V}}\Theta\left( T-\phi(\mathbf{r})\right) N_e^2(\mathbf{r})\, d\mathbf{r}\, , \qquad   \nabla \phi(\mathbf{r})\neq 0 \, ,
\end{equation}

\begin{equation}
    Y_s(T)=\sum_{i} Y_i \, \Theta (T-T_i)\, ,     \qquad        Y_i=\int\limits_{V_i} N_e^2(\mathbf{r})\,d\mathbf{r} \, , \label{Ys-term}
\end{equation}
which are given by the volume integrals over plasma regions ($\widetilde{V}$) and ($V_i$), respectively, with a variable and constant temperature (here, $\Theta(x)$ is the Heaviside step function).

The corresponding density distribution function $y(T)$ (DEM) is thus the sum of the continuous  $y_c(T)$ and singular $y_s(T)$ terms

\begin{equation}
                       dY(T)=y(T)\, dT=\left[ y_c(T)+y_s(T)\right] dT \, , \label{EM}
\end{equation}
which can be defined, respectively, by means of the ``generalized $\delta$-function on the smooth surface'' (Gelfand \& Shilov, 1959)

\begin{equation}
    y_c(T)= \int\limits_{(\widetilde{V})} \delta \left( T-\phi (\mathbf{r})\right) N_e^2(\mathbf{r})\, d\mathbf{r}, \qquad \nabla \phi(\mathbf{r})\neq 0 \, , \label{int_delta}
\end{equation}
which is the volume integral over the $T=\phi(\mathbf{r})$  surface  $\sigma(T)$ disposed in the plasma region $\widetilde{V}$, where the function $\phi(\mathbf{r})$ has non-zero gradient and, as for the volume integrals in the regions $V_i$, where the temperature is constant

\begin{equation}
                      y_s(T)=\sum_i Y_i \, \delta(T-T_i) \, . \label{DEM}
\end{equation}
Equations (\ref{EM}) to (\ref{DEM}) provide the most general invariant (with respect to the system of coordinates) definition of the DEM distribution allowing mathematically correct transformation of the emission power of  the plasma in the volume $V$ (see Eq. (\ref{flux1})) to Eq. (\ref{flux2}). In the case of several surfaces of constant $T$ in the source, the integral in Eq. (\ref{int_delta}) for the function $y_c(T)$ can be presented in the form firstly proposed for smooth function $T(\mathbf{r})$ in Craig \& Brown (1976).
As noted in Urnov et al. (2007b), the presence of singularities (Dirac's delta-function) in the temperature dependence of the DEM distribution is caused by the formal mathematical description of plasma with low spatial temperature gradients in relatively large volumes represented by a ``physical delta-function''. In the limiting case when these gradients are zero in these volumes, the ``physical delta-function'' becomes Dirac's one. In the presence of significant (in terms of volume) regions with a quasi-constant temperature, the smooth function $y_c(T)$ has narrow maxima that transform into singularities $y_s(T_i)$ when passing to the limit of a zero gradient.

The question about the relationship between the contributions of both components, i.e. regions with different temperature gradients, and the total DEM distribution is fairly important for diagnostic purposes.
In the present paper, we consider the emitting plasma model avoiding the approximation of constant temperature in the volume $V$ and omit the singular term in Eq. (\ref{EM}). Below we also use the symbols $\mathbf{S}_{(a)}$   denoting both the summation ($\sum_{(n)}$) or integration ($\int d\nu$), as well as $\delta (a,a')$ denoting Kronecker's symbol or Dirac's $\delta$-function, depending on, respectively, which variables -- discrete ($a=n$) or continuous ($a=\nu$) -- are specified. In our case, the discrete variable is $n=l$ and the continuous one is $\nu = T$. In these notations, Eq. (\ref{flux2}) becomes

\begin{equation}
    F(l)= \mathbf{S}_{(T)}\, G(l,T) y(T)\, , \label{flux3}
\end{equation}
where $F(l)$ is the measured flux in the channel $l$, $G(l,T)$ is the kernel of the integral transformation (\ref{flux3}) and $y(T)$ is the unknown function to be determined.

We refer to as {\it the standard approach}, i.e., the one most often used in astrophysics (which we conventionally call hereafter the ``algebraic'' approach), the spectral inverse problem for the linear system given in Eq. (\ref{flux3}), which can also be written in a matrix form $\mathbf{F}=\mathbf{G}\,\mathbf{y}$, that is assumed to be a special case of the general Fredholm equation of the first kind (see e.g. Craig \& Brown 1976). The solution of this problem is formally given by the matrix $\mathbf{G}^{-1}$ inverse to the matrix $\mathbf{G}$

\begin{eqnarray}
     &  & y(T)  =  \mathbf{S}_{(l)}\, G^{-1}(T,l) F(l)\, , \label{inverse} \\
     &  & \mathbf{S}_{(l)}\, G^{-1}(T,l) G(l,T') =\delta (T,T')\,. \nonumber
\end{eqnarray}
Many numerical techniques based on the algebraic approach have been developed to solve inverse problems. Because of the inherent ill-posedness of the inverse problem based on the integral transformation in Eq. (\ref{flux3}), regularization constraints are used to derive a physically meaningful stable solution $y(T)$ (see e.g. Craig \& Brown 1986).

\noindent{\it In the probabilistic approach}, Eq. (\ref{flux3}) is rewritten by means of the normalized functions

\begin{eqnarray}
    P(l)&=&\frac{F(l)}{\mathbf{S}_{(l')} F(l')}\, , \quad P(l\, |T)=\frac{G(l,T)}{\mathbf{S}_{(l')} G(l',T)}\, , \label{norm_cond}\\
    P(T)&=&\frac{y(T)\, \mathbf{S}_{(l)} G(l,T)}{\mathbf{S}_{(l)} F(l)}\, , \nonumber
\end{eqnarray}
and considered as the formula of the total probability for the distributions $P(l)$, $P(l\, |T)$, and $P(T)$

\begin{equation}
                P(l)= \mathbf{S}_{(T)}\,  P(l\, |T) P(T) \, , \label{flux_norm}
\end{equation}
which are positively determined and satisfy the normalization conditions

\begin{equation}
     \mathbf{S}_{(l)} P(l)=1, \quad \mathbf{S}_{(l)} P(l\,|T)=1 , \quad   \mathbf{S}_{(T)} P(T)=1 \, . \label{norm_cond2}
\end{equation}
These functions can be interpreted as probability distributions for some random variables defined on the field of events $\{l\}$ and $\{T\}$, that satisfy the probability theory postulates: $P(l)$ the probability of a photon being emitted in the spectral channel $l$, $P(T)$ the probability density of a photon being emitted by plasma at the temperature $T$, and $P(l\,|T)$ the conditional probability of the event $l$ at the condition $T$. We note that, in contrast to the algebraic approach, the probabilistic one deals with relative photon fluxes and normalized ``kernels'' of Eq. (\ref{flux_norm}), even if it is considered as a linear integral transformation, and that the unknown ``vector'' $P(T)$ in this equation depends, in addition to the DEM, on the emitting properties $\mathbf{S}_{(l)} G(l,T)$ of the plasma under consideration.

Thus the ill-posed problem arising for some kernels of Eq. (\ref{flux3}) can in principle be well conditioned for the normalized kernels. In addition, the usage of relative fluxes are rather convenient for the analysis of the DEM temperature profiles whenever the spectroscopic data are measured simultaneously by different devices. The formula of the total probability for the distribution $P(T)$ is then given by the expression

\begin{equation}
               P(T) = \mathbf{S}_{(l)} \, P(T|l) P(l) \, , \label{distrib}
\end{equation}
where the a posteriori probability $P(T|l)$, which is connected to the conditional one $P(l\,|T)$ by the Bayesian relation

\begin{equation}
                  P(T|l) = \frac{P(l\,|T)P(T)}{P(l)} \, , \label{Bayes}
\end{equation}
formally provides the inversion of Eq. (\ref{flux_norm}). Multiplying both sides of Eq. (\ref{flux_norm}) by the matrix $P(T|l)$ and summing over $l$, one derives using Eq. (\ref{distrib}) the identity

\begin{eqnarray}
     P(T) & = & \mathbf{S}_{(T')}\,  A(T,T') P(T') =  \nonumber\\
          & = & P(T)\,\, \mathbf{S}_{(l)}  \frac{P(l\, |T) P(l)}{\mathbf{S}_{(T'')} P(l\, |T'')P(T'')} \, , \label{identity}
\end{eqnarray}
where the matrix $A(T,T')$, obtained by means of the Bayesian formula in Eq. (\ref{Bayes}), can be written in the form

\begin{eqnarray}
     A(T,T') & = & \mathbf{S}_{(l)}\, P(T|l) P(l\, |T')= \nonumber\\
             & = &P(T)\,\, \mathbf{S}_{(l)} \frac{P(l\, |T) P(l\, |T')}{\mathbf{S}_{(T'')} P(l\, |T'')P(T'')} \, . \label{matrix}
\end{eqnarray}
As can be seen from Eq. (\ref{matrix}), this matrix depends (nonlinearly) on the ``vector'' $P(T)$ (as well as on the ``inverse'' matrix $P(T|l)$ in Eq. (\ref{distrib})) and thus, in contrast to Eqs. (\ref{flux3}) and (\ref{inverse}), is not a unit matrix in an ordinary algebraic sense. However, the identity in Eq. (\ref{identity}) can be used to formulate the iterative procedure. Interpreting the unknown probability $P(T)$ on the left-hand side of Eq. (\ref{identity}) as step $(n + 1)$ of  the iterative procedure, $P^{(n+1)}(T)$, and on the right-hand side as step $n$, $P^{(n)}(T)$, and taking the known (measured) values $P^{\mathrm{(exp)}}(l)$ for $P(l)$ in Eq. (\ref{flux_norm}), one obtains a recurrence relation for the $P(T)$ temperature profile

\begin{equation}
     P^{(n+1)}(T) = P^{(n)}(T)\,\, \mathbf{S}_{(l)} \frac{P^{\mathrm{(exp)}}(l)}{P^{(n)}(l)} P(l\, |T) \, , \label{iter_formula}
\end{equation}
where $P^{(n)}(l)$ is given by

\begin{equation}
               P^{(n)}(l) = \mathbf{S}_{(T)}\,  P(l\, |T) P^{(n)}(T) \, . \label{int_norm}
\end{equation}
The left-hand side of Eq. (\ref{iter_formula}) may be considered as the estimate of the $n$-th hypothesis, and the correction value $P^{\mathrm{(exp)}}(l)/P^{(n)}(l)$ can be used as a measure of the accuracy of the description of the distribution $P(l)$ over a set of spectral channels $\{ l\}$. To control the rate of the convergence of the iterative procedure in Eq. (\ref{iter_formula}), the estimate based on $\chi^2$ values (a consistency criterion) can be applied (and was actually used in our numerical code along with the ratios $P^{\mathrm{(exp)}}(l)/P^{(n)}(l)$ to check that a sufficient number of iterations had been made to ensure the convergence of the solution)

\begin{equation}
               \chi^2 = \sum_l \frac{\left( I^{\mathrm{(exp)}}(l)-I^{(n)}(l)\right)^2}{I^{(n)}(l)} \, ,
\end{equation}
where $I^{\mathrm{(exp)}}(l)$ and $I^{(n)}(l)=\mathrm{const}\cdot \mathbf{S}_{(T)}  G(l,T) y^{(n)}(T)$ are the observed and calculated photon signals in the channel $l$, respectively. Both of the values, $P^{\mathrm{(exp)}}(l)/P^{(n)}(l)$ and $\chi^2$, also make it possible to reveal the ``incompatible'' spectral lines or broadband channels $\{ l\}$ causing the slow convergence because of systematic errors in the measurements and/or the insufficiency of theoretically predicted luminosity functions $G(l,T)$.

We note that in the absence of these ``incompatible'' channels $\{ l\}$ the solution $P(T_j)$ and, thus, the corresponding DEM $y(T_j)$ from Eq. (\ref{norm_cond}) ($T_j$ are the temperature points in the partition of the whole temperature interval under study, and $j=1,\dots ,N_T$) is the limiting case of the iterative procedure at $n\to\infty$. As seen from Eqs. (\ref{iter_formula}) and (\ref{int_norm}), this solution may be obtained with an arbitrary number $N_T$, if the functions $P(l\, |T)$ are known at any value of its argument $T$. When the chosen temperature grid does not coincide with that for a given temperature response function $G(l,T)$, one should evidently apply either a smoothing procedure or an interpolation. The accuracy of the solution $P(T_j)$ is stipulated by the accuracy of the integrals in Eqs. (\ref{iter_formula}) and (\ref{int_norm}) numerically evaluated with a chosen partition. At the increasing number $N_T$ of temperature points (bins), the function $P(T,N_T)$ converges to its limiting value $P(T)$, giving the most likely solution (i.e., satisfying the ML principle) of the inverse problem within the framework of the probabilistic approach. In contrast to the inversion techniques within the algebraic approach, which are restricted in terms of the choice of the number of temperature bins by the number of channels, with the BIM one should use a temperature grid as fine as needed to derive a smooth and accurate enough solution without producing instabilities related to the ill-posedness of the inverse problem given by Eq. (\ref{flux3}). Thus, in the case of the BIM application the increase in the number of points in the temperature grid increases the accuracy of the solution sought for, if the normalized temperature response functions $P(l\, |T)$ are known at any $T$ and the normalized fluxes $P^{\mathrm{(exp)}}(l)$ are considered as a probability distribution over spectral channels $l$. The analysis of the uncertainties in the solution, which are related to the accuracy of the determination of temperature response functions, is beyond the scope of the present paper and will be presented elsewhere.

To avoid terminological confusion, we underline that in the probabilistic approach the choice of the temperature grid is neither connected nor affected by the temperature resolution of the BIM, if this resolution is
defined as the ability of the method to resolve fine structure in DEM temperature profiles. This means, for example, the ability to reveal closely spaced quasi-isothermal peaks in the DEM employing a given measured spectrum. The minimum temperature interval $\delta T$ between resolved neighboring peaks may be taken as a measure of the ``temperature resolution'' of a method used. In the BIM, the choice of the minimum temperature bin is stipulated by the numerical procedure of evaluation of integrals and by the temperature grid for the given contribution functions, while the temperature resolution of this method depends on the number of channels, the temperature range of their formation, and the temperature sensitivity of channels. At the same time, the absence of any restriction to the choice of the temperature bin is an advantage of the method, which allows us to analyze the temperature resolution of the BIM. An example demonstrating the importance of this analysis can be found in the paper of Landi \& Feldman (2008), where a decrease in the temperature bin from $\Delta\log T = 0.2$ to $\Delta\log T = 0.05$ made it possible to reveal two peaks instead of one (see Fig. 5 in that paper). In contrast to the DEM diagnostic technique used in that paper, the BIM does not require any degree of smoothing over temperature that might introduce artificial effects, as mentioned by its authors. A quantitative analysis of the temperature resolution made on the basis of numerical tests for the spectral sets investigated in the present paper (see Sect. \ref{sect_3}) showed that $\delta T$ is on the order of $\Delta\log T\sim 0.01$ for the line spectra and $\Delta\log T\sim 0.1$ for the broadband data under study.

The zeroth approximation for a priori probability distribution $P(T)$ of the random variable $T$ may be taken in accordance with any theoretical or experimental knowledge. If there is no information about an a priori initial distribution, in accordance with the Bayes' postulate we use a uniform distribution $P^{(0)}(T)=\mathrm{constant}$ for the random variable $T$. One can also show that the normalizing condition and the positivity property of the distribution $P(T)$ are automatically conserved at any step of the iterative procedure in Eq. (\ref{iter_formula}), if only $P^{(0)}(T)\geq 0$ and $\mathbf{S}_{(T)} P^{(0)}(T)=1$.

The iterative procedure developed for normalized variables can be formally carried out over the standard (algebraic) formalization scheme using the definitions in Eq. (\ref{norm_cond}). For the DEM distribution, one obtains

\begin{equation}
 y^{(n+1)}(T)=y^{(n)}(T)\,\, \mathbf{S}_{(l)} \left[\frac{F^{\mathrm{(exp)}}(l)}{F^{(n)}(l)}\right] \left[ \frac{G(l,T)}{\mathbf{S}_{(l')}G(l',T)}\right] \, ,
\end{equation}
where the zeroth approximation is taken in accordance with the a priori probability

\begin{equation}
    y^{(0)}(T) = \frac{P^{(0)}(T)\,\, \mathbf{S}_{(l)}F(l)}{\mathbf{S}_{(l)}G(l,T)} \, . \label{inital_DEM}
\end{equation}

In real applications, the experimental fluxes $F(l)$ in Eq. (\ref{flux3}) are known with some accuracies. To derive the solution of the system of Eqs. (\ref{flux3}), we apply the method of statistical simulation, that is the reconstruction procedure is repeated several times for $M$ different sets of randomly distributed experimental data $F^{\mathrm{(exp)}}(l)$. This simulation finally provides the solution in the form of the mean value

\begin{equation}
    \overline{y(T)}= \frac{1}{M} \sum_{i=1}^M y^{[i]}(T) \label{mean_value}
\end{equation}
and the dispersion

\begin{equation}
    \sigma \left[ y(T)\right] = \sqrt{\frac{1}{M-1} \sum_{i=1}^M \left[ y^{[i]}(T) - \overline{y(T)}  \right]^2 } \, , \label{dispersion}
\end{equation}
which are evaluated over different sets of reconstructed distributions $y^{[i]}(T)$, $i=1,2,\dots,M$ for an arbitrary number of temperature points $N_T$.

This procedure also allows us to estimate the confidence level of the solution thus derived related to statistical errors of measurements for any chosen temperature grid. The calculations performed here in numerical tests and for real observable data showed that the decrease in the temperature bins starting from $\Delta\log T =0.1$ leads to some changes in the DEM confidence level, which, however, are negligible relative to the statistical errors in the measurements for both spectral lines and broadband channels (see also Fig. 4 in Shestov et al. 2010). In the present paper, we found it reasonable to restrict ourselves to a bin size no smaller than $\Delta\log T =0.01$, since the solution satisfying the ML criterion, as well as its confidence level, do not change when the temperature bin continues to decrease. Thus, for such a small bin size the BIM already provides us with smooth and accurate enough solutions of the inverse problem. The quantitative analysis of the accuracy and reliability of the BIM by means of numerical forward-and-back simulations for both sets identified below of spectral channels (see Sect. \ref{sect_3}) and DEM model distributions was the main goal of the present work.

The systematic errors in the values $F^{\mathrm{(exp)}}(l)$ caused by calibration problems have to be determined from the detailed analysis of the intensity fluxes before any restoration procedure can be applied. The BIM application, when used to analyze spectral data from a single instrument, helps us to uncover spectral channels for which experimental and calculated fluxes are ``incompatible'', as mentioned above. On the other hand, the relative values $P(l)$ in Eq. (\ref{norm_cond}) may in principle be defined using in the denominator all the fluxes considered in the analysis, irrespective of the number of instruments used. This means that, when using different instruments, the relative calibration between them can affect $P(l)$ and hence the DEM analysis. We avoid this form of intercalibration by considering the BIM applications to observed data from a single spectroscopic instrument at a time.

We note that the probabilistic approach to the spectroscopic problems under consideration provides a natural formalization of the spectral inverse problem, since it naturally follows from a quantum-mechanical description of the emission process in terms of random variables and probability distributions. At the same time, the algebraic approach, while allowing us to apply effective analytical methods of investigation, needs to be completed because of its intrinsic ill-posed character. A probabilistic treatment of the spectra formation problem represents an alternative mathematical ``language'' of the probability theory providing a natural way to apply statistical methods needed in particular to estimate the confidence level of the solution. It also allows us to define different average temperatures, characterizing various properties of the emitting plasma, as

\begin{equation}
       T_P=\mathbf{S}_{(T)} T P(T)\, , \qquad  T_{EM}=\frac{\mathbf{S}_{(T)} T y(T)}{\mathbf{S}_{(T)} y(T)} \, ,
\end{equation}
where $T_P$ and $T_{EM}$ are the average ``emission power temperature'' for the mean plasma temperature, emitting the main power, and the average ``emission measure temperature'' for the mean temperature of the plasma emission measure, respectively.

At the end of this section, we emphasize the principal difference between the BIM and other DEM methods based on a Bayesian formalism (e.g. Kashyap \& Drake 1998). The methods formulated within the algebraic approach adopting a Bayesian statistical paradigm to analyze the observed spectral data, solve the inverse problem in the standard form of the integral equation (\ref{flux3}). In contrast to these DEM techniques, the BIM is formulated within the framework of the probabilistic approach based on solving Eqs. (\ref{norm_cond})-(\ref{norm_cond2}) using probability distributions. More detailed treatment of the probabilistic approach to the spectral inverse problem will be given elsewhere.

\section{Numerical tests and simulations}
\label{sect_3}

To estimate the efficiency and robustness of our iterative procedure, we carried out some calibration tests and simulations. One of the tests is forward modeling approach (or forward-and-back analysis of artificially generated models) formulated as follows. Using an input model distribution $y_m(T)$ for the DEM (called also parent DEM), we calculate the intensity fluxes in the set of lines or broadband channels considered using Eq. (\ref{flux3}) and the theoretical response functions $G(l,T)$. These fluxes are then assumed to be the ``observed'' ones and used as input data in the DEM reconstruction procedure with the BIM. We thus obtain the calculated distribution $y_c(T)$ and then compare it with the assumed DEM curve, $y_m(T)$.

Another important property of an inverse problem method is its stability. This property is related to the response of the derived distribution, $y_c(T)$, to randomly distributed perturbations in the ``observed'' intensity fluxes. For this purpose, we induced some noise variations in the input data and solved the inverse problem using the BIM.

In our tests, we used three spectral sets of channels $\{l\}$: two sets of spectral lines and one of broadband data. The first set consists of a number of SUMER/SOHO lines reported by Curdt et al. (2004). These lines were used by Landi \& Feldman (2008) for the EM and DEM analysis of an AR spectrum observed by the SUMER instrument onboard SOHO. We used the spectral lines listed in Table 1 of the paper by Landi \& Feldman (2008) excluding only those (given in Table 5 of that paper) that the authors had to discard themselves because of problems related to atomic data physics. Hereinafter, we shall refer to this line set as the ``SUMER line set''. The temperature range covered by the lines from this set spans the interval $\log T \approx 5.6-6.9$.

\begin{figure*}
   \centering
   \includegraphics[width=15cm]{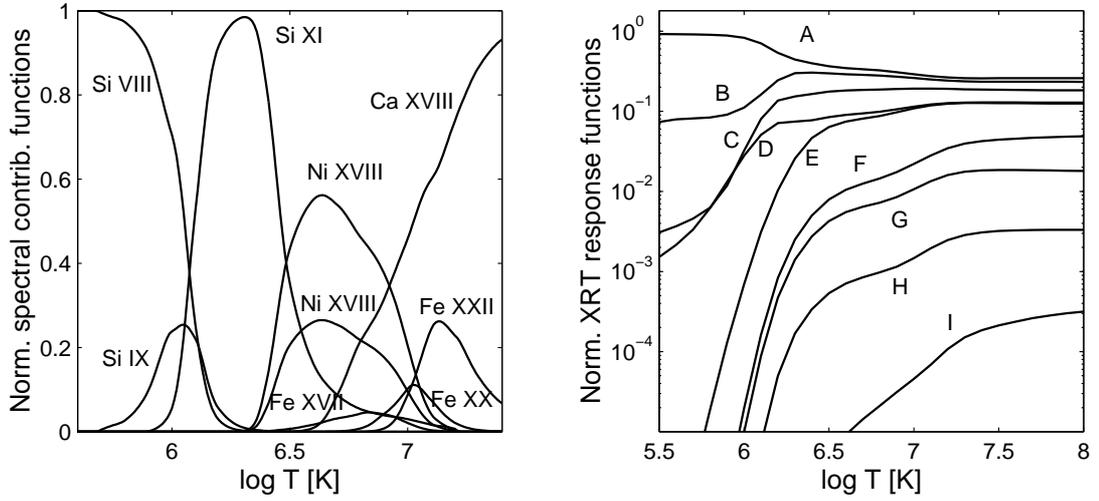}
      \caption{Normalized contribution functions $P(l\, |T)$: the ``SPIRIT line set'' contribution functions (left) and the XRT broadband temperature response functions (right). For the XRT response functions, the logarithmic scale is used. The labels in the right panel indicate the corresponding XRT filters: A = Al\_mesh, B = Al\_poly, C = C\_poly, D = Ti\_poly, E = Be\_thin, F = Be\_med, G = Al\_med, H = Al\_thick, I = Be\_thick.
              }
         \label{FigContrFunc}
   \end{figure*}

The other line set includes the nine spectral lines observed by the SPIRIT spectroheliograph onboard CORONAS-F satellite and used for the DEM analysis of a number of solar ARs in Shestov et al. (2010) (hereinafter referred to as the ``SPIRIT line set''). The lines derived from EUV spectra in the wavelength range 280--330 \AA\ are Ni XVIII 291.98 \AA, Fe XXII 292.46 \AA, Si IX 296.11 \AA, Ca XVIII 302.19 \AA, Si XI 303.33 \AA, Fe XX 309.29 \AA, Si VIII 319.84 \AA, Ni XVIII 320.57 \AA, and Fe XVII 323.65 \AA\ (see also Fig. \ref{FigContrFunc} for the corresponding normalized contribution functions). For this set, the temperature coverage constitutes $\log T \approx 5.9-7.2$. To be consistent with Landi \& Feldman (2008) and Shestov et al. (2010) in our DEM analysis, we used the same version of the CHIANTI database (version 5.2.1 -- Dere et al. 1997; Landi et al. 2006) to calculate line emissivities, as well as the same ion abundances (Mazzotta et al. 1998) and the coronal element abundances (Feldman et al. 1992).

In the case of broadband spectroscopy, we chose the nine solar coronal Hinode/XRT filter channels (Golub et al. 2007). The XRT provides high resolution images of the emitting coronal plasma in the temperature range $6.1 < \log T < 7.5$. One of the main differences between the line and the broadband spectroscopy consists in the functional behavior of the temperature response functions $G(l,T)$. The spectral line contribution functions have prominent temperature dependence and are well separated from each other, i.e. one can say that they are linearly independent. In contrast, as the broadband temperature response functions contain many spectral lines and the continuum, they are more smooth and may have similar forms in different channels. This is why the information from the DEM analysis in the case of broadband data is more limited and less detailed.
The different temperature behavior of the response functions in the case of narrow (SPIRIT) and broadband (XRT) instruments is shown in the left and right plots, respectively, of Fig. \ref{FigContrFunc}. Here these functions are plotted as normalized ones, $P(l\,|T)$, versus the logarithm of temperature. For the sake of visualization, the XRT response functions are also presented on a logarithmic scale.
The reason for this difference is that the effective areas of the broadband filters partially overlap and the corresponding spectral channels are not completely independent.

\subsection{Artificial DEMs reconstructed to compare the three spectral sets}\label{sec_spectra}

\begin{figure*}
   \centering
   \includegraphics[width=17cm]{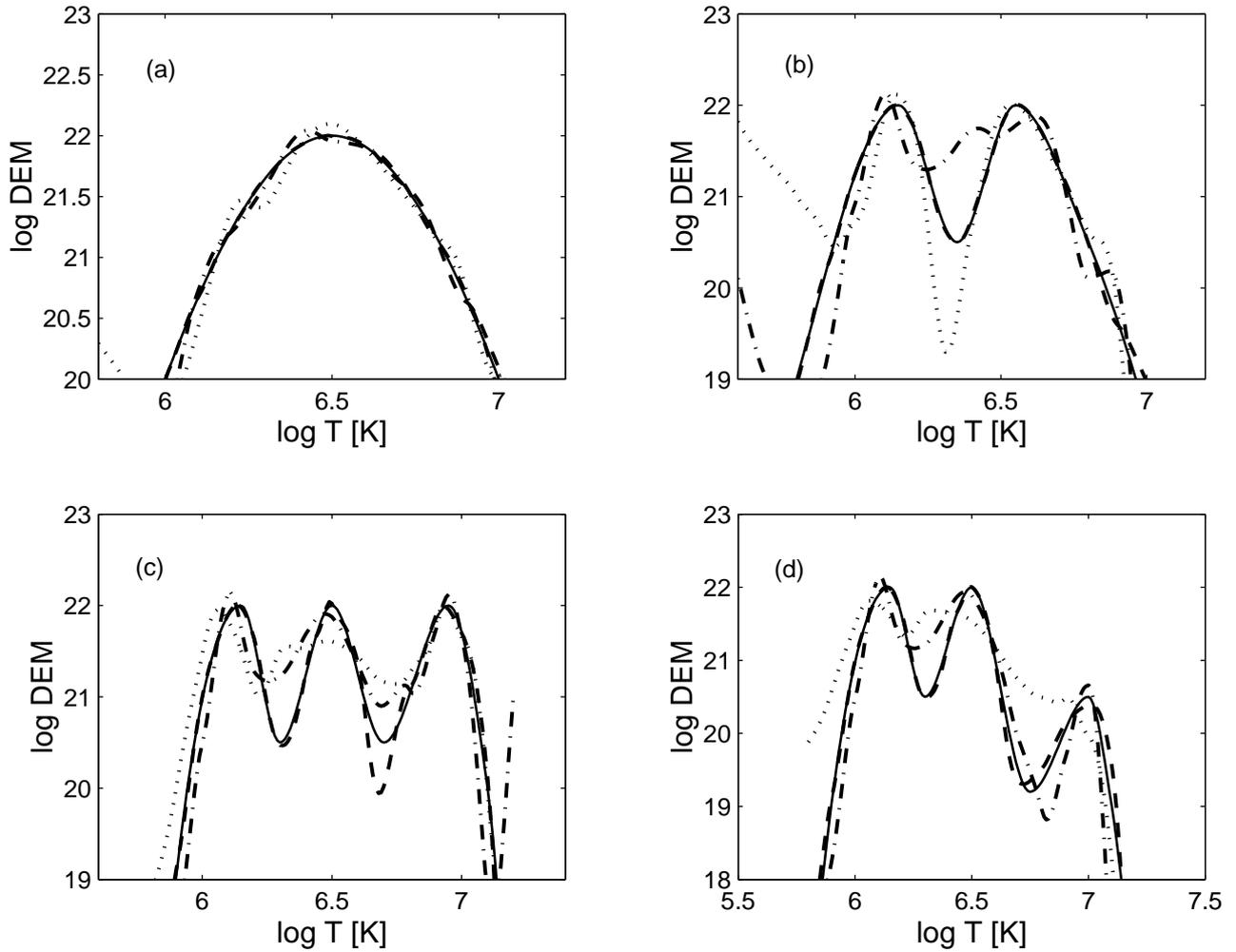}
      \caption{Results of the forward modeling tests. The model DEM distributions (solid lines) are compared with the reconstructed ones for three sets of spectral data: the SUMER line set (dashed lines), the SPIRIT line set (dash-dot lines), and the nine XRT filter channels (dotted lines). The {\it y} axis gives the logarithm of DEM in arbitrary units.
              }
         \label{FigCompDEM}
   \end{figure*}

We started by fitting different assumed model DEMs with one, two, and three characteristic peaks. We included a particular case where the high temperature peak was varied in amplitude until it was far weaker than the others (see Fig. \ref{FigCompDEM}(d)). This choice was motivated by the results of the DEM and EM calculations in active regions (see e.g. Urnov et al. 2007b; Landi \& Feldman 2008; Reale et al. 2009).

The model DEM curves were convoluted with the kernels $G(l,T)$ for the channels $l$ in each spectral set to obtain the corresponding artificial fluxes. The BIM algorithm was then applied to reconstruct the distributions $y_c(T)$. The results of these calculations are presented in Fig. \ref{FigCompDEM}. In all the plots (a)--(d), the solid line corresponds to the assumed DEM curves, $y_m(T)$. A comparison shows that for all the curves the assumed and calculated DEMs derived using the SUMER line set (dashed lines) have practically no differences. Between the SPIRIT line set (dash-dot lines) and XRT broadband channels (dotted lines), some differences are found, but qualitatively the reconstructed DEMs reproduce the assumed distributions with good accuracy, in particular the magnitudes and positions of the DEM distribution maxima.

We also note that on average for the SPIRIT line set the agreement between the reconstructed and model distributions is closer than for the XRT channels. This is because of the above-mentioned functional difference between the normalized contribution functions $P(l\,|T)$ of spectral lines and broadband channels (see Fig. \ref{FigContrFunc}).

\begin{figure*}
   \centering
  \includegraphics[width=15cm]{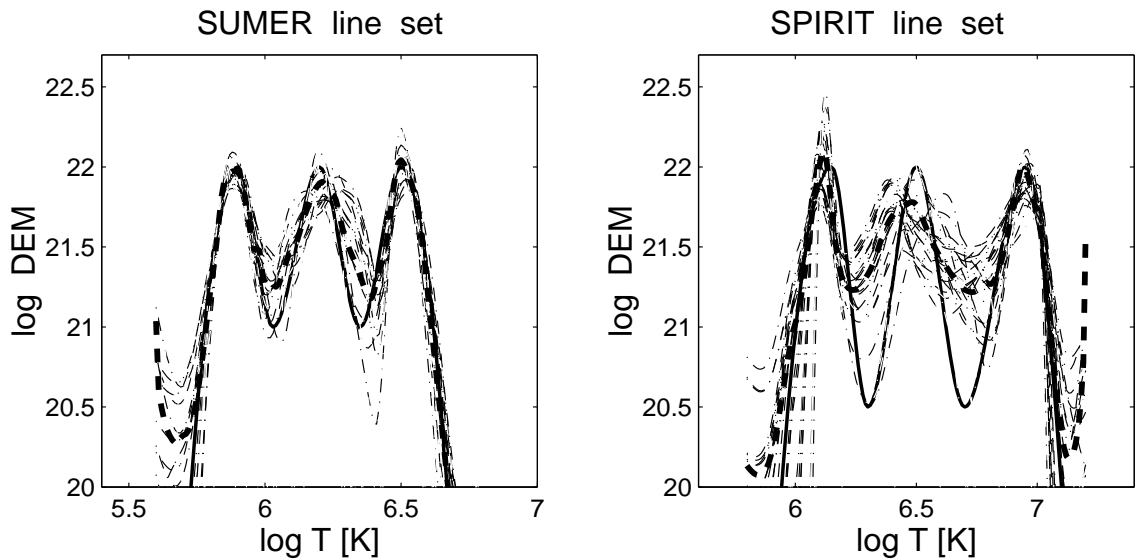}
      \caption{Stability tests of the BIM. The heavy solid line stands for the model DEMs; the dash-dot lines indicate the calculated distributions using the perturbed line fluxes; the heavy dashed line indicates the median fit. The {\it y} axis gives the logarithm of DEM in arbitrary units.
              }
         \label{FigStabDEM}
   \end{figure*}

To demonstrate the stability of the BIM, we also performed corresponding tests using the SUMER and SPIRIT line sets. We used two model distributions with three characteristic peaks (heavy solid lines in Fig. \ref{FigStabDEM}) in the temperature intervals $\log T \approx 5.6-6.8$ for the SUMER line set (left), and $\log T \approx 5.8-7.2$ for the SPIRIT line set (right). The choice of the temperature intervals was guided by the different temperature sensitivity of the two instruments SUMER and SPIRIT.
Using the assumed input DEM distributions, we first calculated the corresponding model fluxes, then we perturbed them randomly with amplitudes of 20\%. We then used the randomly perturbed fluxes as input to reconstruct ``perturbed distributions'' (dash-dot lines). The heavy dashed lines indicate the optimal mean curves. The quantitative outcome of the stability tests shows that the induced perturbations result in the compatible disturbances in the calculated DEM distributions.

\subsection{Tests on different number of XRT channels}

We now investigate in more detail the capabilities of the BIM when applied to the wide band XRT data. In Fig. \ref{FigCompDEM}, we showed that the use of the full set of the nine filters is enough for the identification of up to three peaks.
Here we apply the forward modeling described above to both all nine and only five XRT filters to test the diagnostic ability of the BIM technique under different data conditions. In particular, we aim to test the BIM under the conditions of   Reale et al. (2009), where only five XRT filters were used (Al\_poly, C\_poly, Be\_thin, Be\_med, Al\_med) for an EM analysis of an AR. In Sect. \ref{sect_43}, we apply the BIM to the same data. In the analysis of XRT broadband data, we use the same bin size as adopted for the filter XRT response functions, i.e. $\Delta\log T = 0.1$, which happened to be similar to the value $\delta T$ derived from temperature resolution tests for the XRT channels.

\begin{figure*}
   \centering
   \includegraphics[width=14cm]{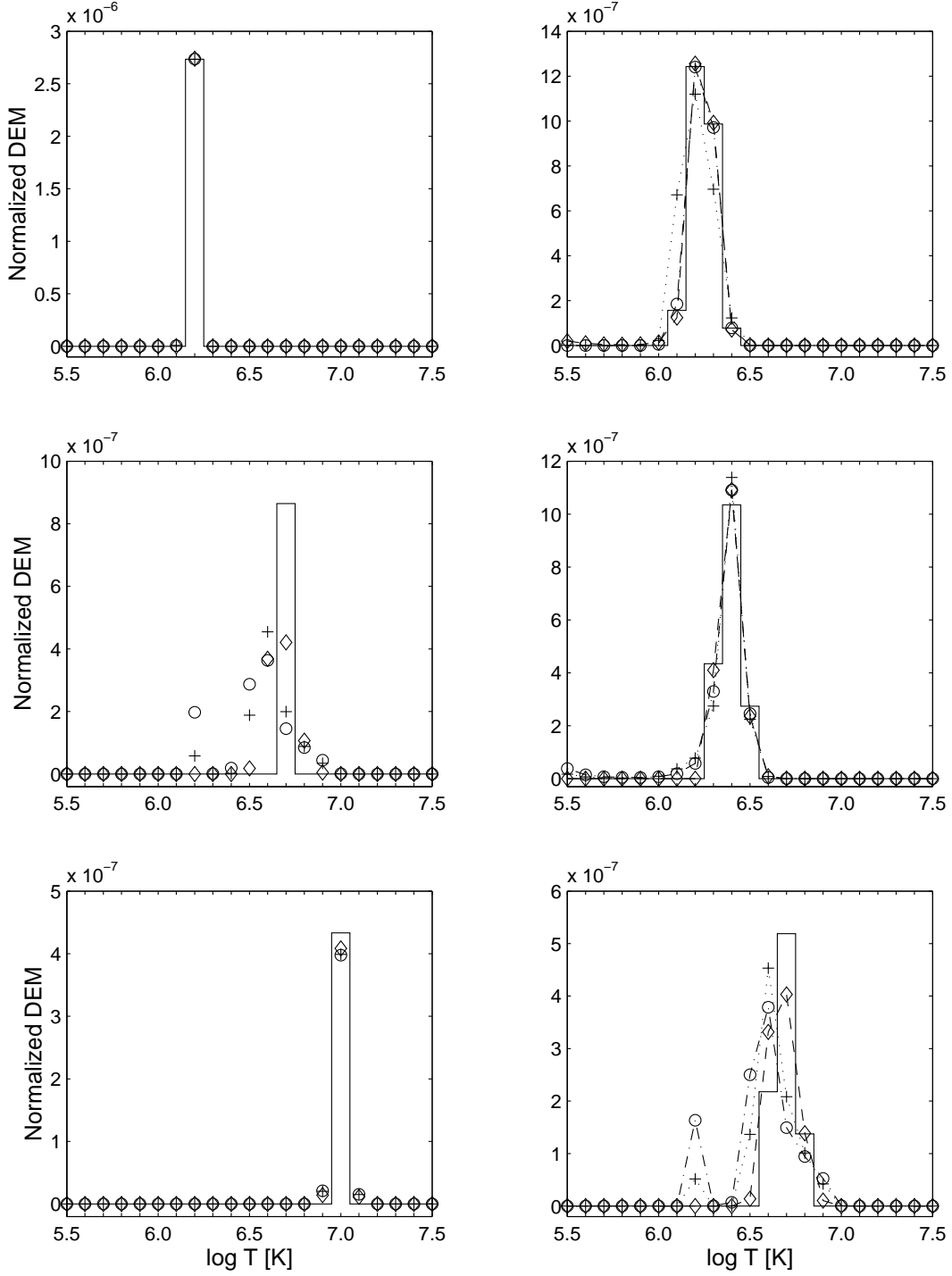}
      \caption{Forward modeling tests for DEM reconstruction with XRT channels. In the left column, the fits of single-temperature component DEMs are shown centered at $\log T = 6.2, 6.7, 7.0$ and with the width $\Delta\log T = 0.1$. In the right column, the fittings for several multi-temperature DEMs are given. The model DEMs (solid lines) are compared with the reconstructed ones for three groups of XRT channels: the diamonds (left column) and diamonds with dashed line (right column) indicate the solutions for the nine XRT filters with standard calibration; the pluses and pluses with dotted line represent the five XRT filters with standard calibration; the circles and circles with dash-dot line stand for the five XRT filters with calibration of Reale et al. (2009). The DEMs are given as normalized distributions $\rho (T)$ in units of [K$^{-1}$].
              }
         \label{Fig3}
   \end{figure*}

To calculate the synthetic fluxes in the case of five channels, we adopted both the standard calibration and a new one developed by Reale et al. (2009) to identify any dissimilarity.
The first series of tests were made for the simplest case where the DEM distribution is isothermal and of width $\Delta \log T=0.1$. We made the position of the peak vary from  $\log T=6.0$ to 7.0. Figure \ref{Fig3}, first column, reports, as an example, the results for $\log T=6.2$, 6.7, and 7.0. The plots show DEM distributions as functions of the logarithm of temperature.
The DEM curves are given as functions $\rho (T)=y(T)/\int y(T)\, dT$ (see e.g. Golub et al. 2007, Schmelz et al. 2009a) normalized to unity in a temperature range under study.
In the plots, the solid line is the parent DEM, while the diamonds and plus curves are the reconstructed DEMs using, respectively, nine and five filters with the standard calibration. The curve with circle symbols refers to the derived DEM with five filters assuming the new calibration.

The results of the tests are similar for both calibrations with the plasma temperatures that can be clearly identified, except the temperature interval $\log T\approx 6.7-6.8$. In this temperature range, the reconstructed DEMs are wider than the model ones, and in the case of five XRT channels there is a shift towards lower temperatures, as shown in the middle left panel of Fig. \ref{Fig3}. Moreover, when using the five channels to reconstruct DEMs at $\log T\approx 6.7-6.8$, some artifacts for $\log T\approx 6.2$ such as smaller plasma components and differences between the two calibrations seen in Fig. \ref{Fig3} appear. The cause of this behavior might be the flat temperature responses of the XRT channels at $\log T\approx 6.7-6.8$, leading to a poor temperature discrimination (see Fig. \ref{FigContrFunc}, the right plot). In general, these results illustrate the possible ways in which the BIM can be applied to the XRT data. In spite of their wide temperature responses, five filters are enough not only to identify but to quantify an isothermal plasma (as was also noted by Reale et al. 2009) at temperatures $\log T < 6.7$ and $\log T > 6.8$, and have some quantitative diagnostic limitations at $\log T\approx 6.7-6.8$.

A second run of tests was performed to consider wider DEM distributions in the same temperature range. In the first step we performed simulations with one-component DEMs and found similar results to the previous case: two sets of five filters with both the standard and new calibrations provide similar DEM results to nine filters with the standard calibration (right column in Fig. \ref{Fig3}). There is also a shift towards lower temperatures, such that the cool plasma artifacts at $\log T\approx 6.2$ are found at $\log T\approx 6.7-6.8$ when only five filters are used (bottom right panel of Fig. \ref{Fig3}).

\begin{figure*}
   \centering
   \includegraphics[width=14cm]{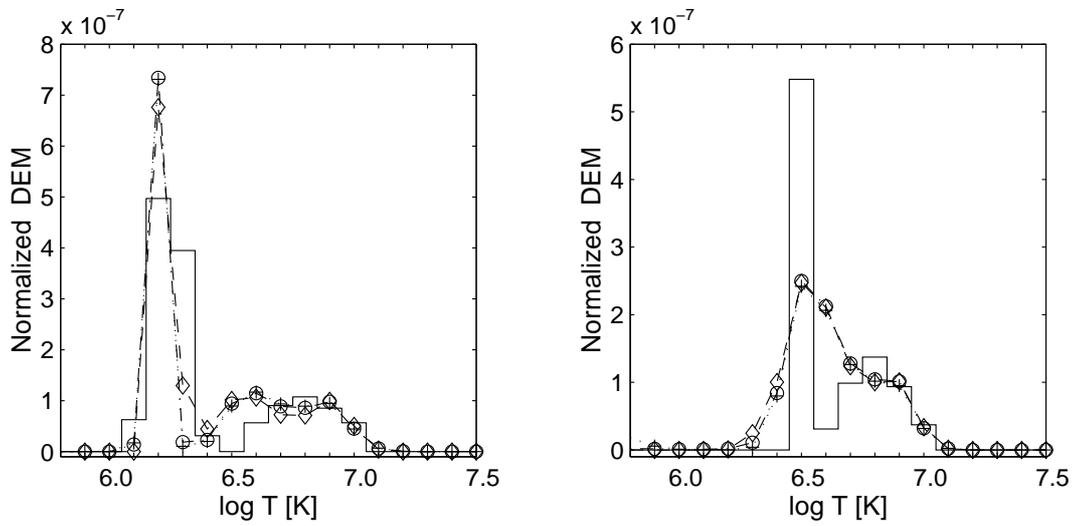}
      \caption{Same as Figure \ref{Fig3}, but the model DEMs have both a cool component and a weaker hot component.
              }
         \label{Fig4}
   \end{figure*}

An additional step was included where we modeled a two component plasma consisting of a main cool component and a weaker hot component (see Fig. \ref{Fig4}). This choice was motivated by the data analysis results for active regions described in Sect. \ref{intro}. The ability of the inversion method to detect hot plasma in the presence of the main cooler bulk component is of primary importance to the study of the coronal heating problem.

The peak of the main component was made to vary between $\log T=6.2$ and 6.5, while the minor component was varied between $\log T=6.4$ and 7.0. The widths were also made to vary as in previous cases. Some of these tests are similar to that of Reale et al. (2009).

Considering the wide temperature responses of these filters, the BIM continues to operate well in the presence of  this secondary component. When the main component is cool enough to be isolated from the secondary (left plot in Fig. \ref{Fig4}), the BIM still resolves them. In the case of the secondary component at $\log T=6.8$, some artifacts appear. The position of the main peak is clearly identifiable, but the height and width are not completely reproduced. We integrated the amount of plasma in the temperature range $\log T = 6.05-6.55$ (i.e. the bins at the points $\log T = 6.1-6.5$) and found that the EM determined from the BIM reconstruction process agrees with the cool component model EM at the limits of a few percent. The hotter component is clearly detected even though a cooler wing is introduced. No clear difference is found among the choice of filters, even though we may say that the nine filters allow the best solution.

The opposite effect is found for the cool component when the two components are close to each other (right plot). The whole DEM is smoothed and widened, even though the position of the cool peak maximum and the hotter part of the second hump can be clearly identified. Again, no difference is found among the different choice of filters. The integration of the DEM over the temperature range $\log T = 6.35-6.65$ (i.e. the bins at the points $\log T = 6.4-6.6$) showed that the corresponding EM coincides with the hot component model EM to an accuracy of a few percent.

In conclusion to this section, we have shown that despite some minor smoothing affecting our results, our tests are generally positive. Some systematic differences are identifiable as artificial cool components, which may be caused by a combination of the characteristics of the XRT filters and the BIM. This is particularly visible when five channels only are used.
To verify this last result, we also carried out forward modeling tests for the SPIRIT line set in the case of single temperature distributions as discussed for the XRT case. These calculations have shown that in the temperature range $\log T \approx 6.0-7.0$ this spectral set reproduces single temperature DEMs to very high accuracy and clearly determined positions of temperature maxima.

\section{Application to observable data and comparison with other DEM analysis results}\label{sect_4}

\subsection{SUMER/SOHO AR line spectrum}

The BIM approach presented in Sect. \ref{sect_2} was applied to the analysis of an AR line spectrum observed by the SUMER/SOHO instrument and analyzed by means of EM and DEM diagnostic techniques in Landi \& Feldman (2008). In our analysis, we used the same set of line intensities from Table 1 of Landi \& Feldman (2008).

The calculations are performed using the method of statistical simulation. First of all, a set of values $F(l)$ on the left side of Eq. (\ref{flux3}) is generated by randomly varying the observed fluxes in each line $l$ from the ``SUMER line set'' with some induced uncertainties. The corresponding distribution functions of these variations are assumed to be Gaussian with centroids equal to the measured fluxes and widths equal to uncertainties that were assumed to be about 10--30\% for different lines (Curdt et al. 2004). Using this data set, the BIM is applied to reconstruct a DEM($T$) distribution. This process is repeated and we finally derive from this statistical procedure the optimal mean DEM curve and its confidence level (see Eqs. (\ref{mean_value}), (\ref{dispersion}) in Sect. \ref{sect_2}).

\begin{table*}
\caption{List of spectral lines from the ``SUMER line set'' and ratios of their calculated to observed intensities calculated by means of the BIM.}
\label{table:1}
\centering
\begin{tabular}{l r c c|l r c c}
\hline\hline
Ion & $\mathrm{\lambda}$\, , \AA & $\log T_{\mathrm{max}}$ & Calc. /\, Obs. & Ion & $\mathrm{\lambda}$\, , \AA & $\log T_{\mathrm{max}}$ & Calc. /\, Obs.  \\
\hline
Ne VII & 895.16 & 5.7 &  $1.47\pm 0.32\, (1.14)$ & S X & 787.43 & 6.2 & $0.86\pm 0.17\, (0.94)$  \\
Ne IX & 1248.08 & 6.2 & $0.94\pm 0.19\, (0.90)$ & S XI & 552.09 & 6.3 & $0.80\pm 0.24\, (1.20)$ \\
Mg VII & 868.08 & 5.8 & $0.67\pm 0.14\, (0.67)$ & S XI & 574.88 & 6.3 & $0.76\pm 0.16\, (1.16)$ \\
Mg VIII & 762.65 & 5.9 & $1.02\pm 0.31\, (1.09)$ & Ar IX & 642.53 & 5.8 & $0.77\pm 0.23\, (0.76)$ \\
Mg VIII & 769.38 & 5.9 & $1.06\pm 0.32\, (1.13)$ & Ar XI & 745.80 & 6.3 & $0.91\pm 0.27\, (1.20)$ \\
Mg VIII & 772.29 & 5.9 & $1.10\pm 0.23\, (1.18)$ & Ar XI & 1392.10 & 6.3 & $0.88\pm 0.17\, (1.14)$ \\
Mg VIII & 782.37 & 5.9 & $1.02\pm 0.21\, (1.09)$ & Ar XII & 670.31 & 6.4 & $0.96\pm 0.29\, (1.10)$ \\
Mg VIII & 789.44 & 5.9 & $0.70\pm 0.14\, (0.75)$ & Ar XII & 1018.75 & 6.4 & $1.13\pm 0.23\, (1.29)$ \\
Mg IX & 706.04 & 6.0 &   $1.16\pm 0.12\, (1.03)$ & Ar XII & 1054.59 & 6.4 & $1.06\pm 0.21\, (1.21)$ \\
Mg IX & 749.56 & 6.0 &   $1.18\pm 0.24\, (1.04)$ & Ar XIII & 656.69 & 6.5 & $0.76\pm 0.23\, (0.74)$ \\
Mg XI & 1043.28 & 6.5 &  $0.79\pm 0.24\, (0.78)$ & Ar XIII & 1330.53 & 6.5 & $0.73\pm 0.22\, (0.71)$ \\
Al VIII & 1057.86 & 6.0 & $0.82\pm 0.25\, (0.72)$ & Ca IX & 821.22 & 5.9 & $0.73\pm 0.22\, (0.67)$ \\
Al IX & 691.54 & 6.0 &  $0.82\pm 0.25\, (0.62)$ & Ca XIV & 579.85 & 6.5 & $0.78\pm 0.15\, (0.72)$ \\
Al IX & 703.64 & 6.0 &  $0.80\pm 0.24\, (0.61)$ & Ca XIV & 880.40 & 6.5 & $0.89\pm 0.09\, (0.82)$ \\
Al X & 670.02 & 6.1 &   $0.85\pm 0.25\, (0.75)$ & Ca XIV & 943.63 & 6.5 & $1.00\pm 0.10\, (0.92)$ \\
Si VII & 1049.26 & 5.8 & $1.00\pm 0.30\, (0.93)$ & Ca XV & 1098.48 & 6.6 & $0.89\pm 0.18\, (0.81)$ \\
Si IX & 676.51 & 6.0 &  $1.09\pm 0.22\, (0.89)$ & Fe VIII & 721.26 & 5.6 & $0.84\pm 0.18\, (0.69)$ \\
Si IX & 694.69 & 6.0 &  $0.86\pm 0.09\, (0.70)$ & Fe X & 1028.04 & 6.0 &  $0.84\pm 0.17\, (0.68)$ \\
Si IX & 950.16 & 6.0 &  $1.15\pm 0.11\, (0.94)$ & Fe XI & 1467.07 & 6.1 & $1.14\pm 0.11\, (0.92)$ \\
Si X & 624.71 & 6.1 &   $0.90\pm 0.09\, (0.87)$ & Fe XII & 1242.00 & 6.1 & $0.95\pm 0.09\, (0.90)$ \\
Si X & 638.92 & 6.1 &   $1.04\pm 0.10\, (1.02)$ & Fe XII & 1349.37 & 6.1 & $0.88\pm 0.09\, (0.83)$ \\
Si X & 649.21 & 6.1 &  $0.82\pm 0.16\, (0.80)$ & Fe XVII & 1153.16 & 6.6 & $1.10\pm 0.11\, (1.07)$ \\
Si XI & 564.00 & 6.2 & $0.84\pm 0.17\, (1.11)$ & Fe XVIII & 974.84 & 6.8 & $1.22\pm 0.13\, (1.24)$ \\
Si XI & 580.91 & 6.2 & $1.07\pm 0.11\, (1.41)$ & Fe XIX & 1118.07 & 6.9 & $1.20\pm 0.40\, (1.28)$ \\
S X & 776.25 & 6.2 &  $1.14\pm 0.23\, (1.26)$ & Ni XV & 1033.01 & 6.4 & $0.86\pm 0.18\, (1.24)$ \\
\hline
\end{tabular}
\tablefoot{
$T_{\mathrm{max}}$ is the temperature of the maximum abundance for each ion (Mazzotta et al. 1998). In brackets the intensity ratios obtained with the DEM distribution of Landi \& Feldman (2008) are given.\\
}
\end{table*}

\begin{figure*}
   \centering
   \includegraphics[width=9cm]{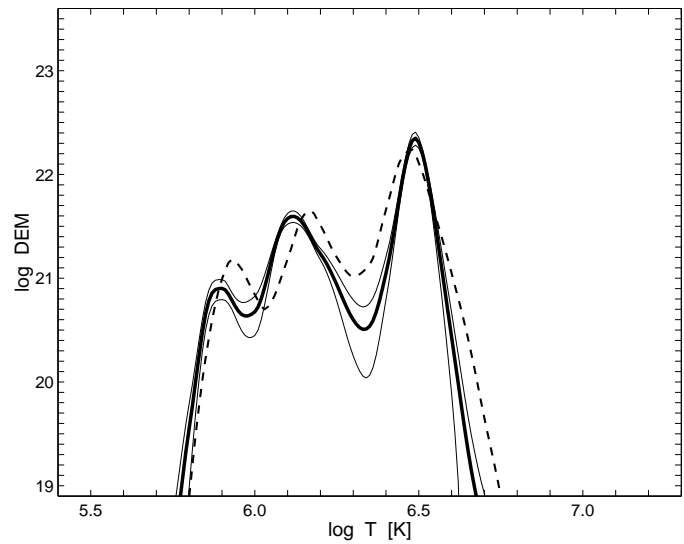}
      \caption{Results of the DEM analysis for the SUMER/SOHO line spectrum. The heavy solid line represents the optimal median curve; the thin solid lines indicate the limits of confidence level for the DEM solution; the heavy dashed line is the data curve by Landi \& Feldman (2008). The DEM is in units of [$\mathrm{cm}^{-5}\cdot\mathrm{K}^{-1}$].
              }
         \label{FigDEMSUMER}
   \end{figure*}

The final optimal mean DEM curve is plotted against the logarithm of temperature in Fig. \ref{FigDEMSUMER} (heavy solid line). The confidence level of this solution (thin solid lines) was obtained by means of statistical simulation. We compare our results with those (dashed line) of Landi \& Feldman (2008), where these authors applied the Landi \& Landini (1997) iterative DEM diagnostic technique. The comparison shows the very good agreement between the DEMs obtained by both techniques. However, there are some quantitative differences. In Fig. \ref{FigDEMSUMER}, one can discern the three characteristic peaks of the DEM distribution. For the curve by Landi \& Feldman (2008), these peaks are located at the temperatures $\log T \approx 5.93$, $\log T \approx 6.17$, and $\log T \approx 6.48$. The position of our higher temperature peak at $\log T \approx 6.48$ coincides with the Landi \& Feldman's calculations in terms of both temperature and DEM values very well. Our middle hump is slightly shifted in temperature to the left ($\log T \approx 6.12$) and has DEM values that are close to data in the aforementioned paper. The cool peak is also shifted to the left ($\log T \approx 5.88$) and its maximum has a smaller DEM value by a factor of $\sim$\, 1.5.

In our study of SUMER/SOHO data, we also identified lines for which the ratios of calculated to observed fluxes are in considerable disagreements (see Sect. \ref{sect_2} for the ``incompatible'' flux ratios). We assume that these lines have some defects that are perhaps related to atomic physics problems, chemical composition of coronal plasma, or incorrect line identifications etc. The ``good'' lines, i.e. the ``compatible'' ones of good accuracy, are presented in Table \ref{table:1} where their calculated-to-observed flux ratios are also listed.

To estimate and compare the accuracy of the results of Landi \& Feldman (2008), we calculated the fluxes in the spectral lines using their DEM curve (dashed line in Fig. \ref{FigDEMSUMER}). The corresponding calculated-to-observed flux ratios are placed in brackets in Table \ref{table:1}. Using the spectral data from Table \ref{table:1}, we also evaluated $\chi^2$ values for both the BIM ($\chi^2\mathrm{(BIM)}$) and Landi \& Feldman ($\chi^2\mathrm{(Landi)}$) results and found that $\chi^2\mathrm{(BIM)}/\chi^2\mathrm{(Landi)}\approx 0.6$. In any case, one may say that our analysis qualitatively confirms both the calculations and main conclusions of Landi \& Feldman (2008).

\subsection{SPIRIT/CORONAS-F AR line spectra data}\label{sect_42}

We describe our validation of the results of Shestov et al. (2010). The analysis was performed for a series of ARs observed by the multichannel RES spectroheliograph in the SPIRIT experiment onboard the CORONAS-F satellite.

The SPIRIT instrumentation recorded simultaneously monochromatic full-Sun images in EUV lines and the X-ray Mg XII 8.42 \AA\ resonance line. The RES spectroheliograph in the Mg XII channel can only register the emission of hot plasma with temperatures $T\gtrsim 5$ MK, and at the same time the luminosity function maximum of the Mg XII 8.42 \AA\ resonance line occurs at $\approx$ 10 MK. This implies that the monochromatic full-Sun Mg XII 8.42 \AA\ images provide direct confirmation of the presence of a hot plasma.

\begin{figure*}
   \centering
    \includegraphics[width=9cm]{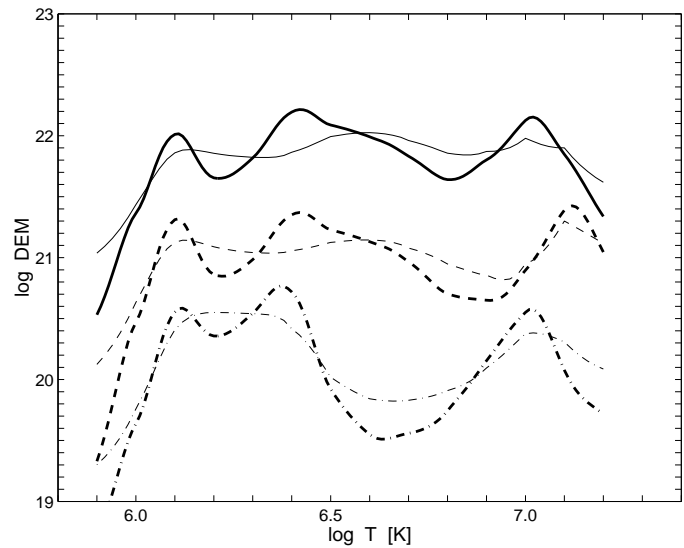}
      \caption{Example of DEM temperature profiles reconstructed with the BIM (thick curves) for the three SPIRIT/CORONAS-F line spectra ARs and the comparison with the corresponding results (thin curves) of Shestov et al. (2010): NOAA 9742, 28.12.2001 at 21:21UT (solid lines); NOAA 9906, 20.04.2002 at 19:24UT (dashed lines); NOAA 0223, 27.12.2002 at 09:06UT (dash-dot lines). The DEM is in arbitrary units.
              }
         \label{FigDEMSPIRIT}
   \end{figure*}

Using a set of EUV lines in the 280--330 \AA\ spectral band (called ``SPIRIT line set'' in the present work), Shestov et al. (2010) evaluated DEM temperature distributions for twelve AR objects. The results obtained there allowed us to infer the existence of non-flaring ARs with different temperature contents and the presence of hot plasmas with temperatures $\log T\sim 6.8-7.2$. We revised the same data and performed the DEM analysis with our BIM algorithm. Figure \ref{FigDEMSPIRIT} shows an example of the DEM temperature profiles for three ARs (objects no. 2, 4, and 7 in Table 1 of Shestov et al. 2010). In Table \ref{table:2}, we provide the observed to calculated flux ratios for the ``SPIRIT line set'' derived from the DEM distributions in Fig. \ref{FigDEMSPIRIT}. They are compared with the intensity ratios from Shestov et al. (2010) (in brackets).

As can be seen from Fig. \ref{FigDEMSPIRIT}, our DEM distributions consist of three prominent plasma components: the two plasma structures similar to quiet-Sun (at $\log T\approx 6.1$) and active region (at $\log T\approx 6.4$) plasmas, and a hot plasma component at $\log T\approx 7.0$. Figure \ref{FigDEMSPIRIT} also indicates that in non-flaring ARs the EM of hot coronal plasma at temperatures $\sim$ 9--15 MK may be comparable to the plasma EM at $\sim$ 2--4 MK.

Our results qualitatively confirm those of Shestov et al. (2010), in particular the main conclusion of that paper about the relative quantity of hot plasma in ARs remains unchanged. However, Figure \ref{FigDEMSPIRIT} reveals some quantitative differences regarding the positions of plasma components and their DEM magnitudes. In contrast to the positions of hot and cool components, which agree relatively well, there are some significant dissimilarities between the middle temperature humps (the top and middle curves in Fig. \ref{FigDEMSPIRIT}). Our detailed analysis showed that differences in the magnitudes of the DEM peaks are mainly caused by our larger number of iteration steps (50-100 iterations versus 5-10 ones made in Shestov et al. (2010) to obtain smoothed DEM curves) and, to a much lesser extent, with the interpolation procedure (the smooth spline interpolation versus the simple linear one in Shestov et al. 2010) for contribution functions and DEM solutions (see Sect. \ref{sect_2}).

Finally, we have found that the shift in middle temperature is stipulated by the substantially different choice of the zeroth approximation for the DEM profile $y^{(0)}(T)$ (see Eq. (\ref{inital_DEM})) for an a priori distribution, $P^{(0)}(T)=1$ in our calculations versus $y^{(0)}(T)=\mathrm{constant}$ in Shestov et al. (2010), adopted also as the default choice in the CHIANTI iterative procedure chianti\_dem.pro. If the BIM is provided with a priori distribution $P^{(0)}(T)$ corresponding to the $y^{(0)}(T)=\mathrm{constant}$, it reproduces Shestov et al. (2010) results for the position of the middle temperature peak. We note that for a large number of lines (i.e. the SUMER/SOHO data) the solution does not depend on $P^{(0)}(T)$, while in the case of their small number (i.e. the SPIRIT/CORONAS-F data), the choice of the zeroth approximation may produce noticeably different solutions. This result demonstrates the importance of an a priori initial DEM distribution in the context of the probabilistic approach, which in our studies is justified by the Bayes' postulate (``equal lack of knowledge''). The forward-and-back simulations made in the present paper have confirmed that this choice of a priori distribution instead of an ``initial guess'' $y^{(0)}(T)=\mathrm{constant}$, provides a more reliable solution resembling with a rather high accuracy the shape of the peaks in the model DEM distributions as well as their temperature positions (within $\Delta\log T\approx 0.05$) even with nine spectral lines (see Fig. \ref{FigCompDEM} for the SPIRIT line set). It has also been found that the increase in this number of lines (in the case of the SUMER line set) by a factor of about three provides a solution that practically coincides with the model distribution and becomes independent of a priori DEM distribution. This solution provides an accuracy of order $\Delta\log T\sim 0.01$, which is close to the temperature resolution $\delta T$ derived from numerical tests for the SPIRIT and SUMER line sets.

\begin{table}
\caption{Ratios of the observed to calculated intensities for the ``SPIRIT line set'' derived by means of the BIM.}
\label{table:2}
\centering
\begin{tabular}{l c c c c}
\hline\hline
Ion & $\mathrm{\lambda}$\, , \AA &  Obs. /\, Calc. & Obs. /\, Calc. & Obs. /\, Calc.   \\
    &                            &   NOAA 9742 & NOAA 9906 & NOAA 0223 \\
\hline
Ni XVIII & 291.98 &  0.88 (0.88) & 1.01 (1.03) & 0.91 (0.90)  \\
Fe XXII  & 292.46 &  1.04 (1.21) & 1.18 (1.57) & 1.06 (0.91)  \\
Si IX    & 296.11 &  1.10 (1.31) & 1.15 (1.39) & 1.11 (1.28)  \\
Ca XVIII & 302.19 &  0.93 (0.86) & 0.93 (0.83) & 0.91 (0.92)  \\
Si XI    & 303.33 &  1.00 (0.99) & 0.99 (0.98) & 1.00 (0.99)  \\
Fe XX    & 309.29 &  1.34 (1.68) & --          & 1.30 (1.71)  \\
Si VIII  & 319.84 &  0.94 (0.88) & 0.91 (0.84) & 0.93 (0.87)  \\
Ni XVIII & 320.57 &  1.30 (1.32) & 1.00 (1.02) & 1.15 (1.14)  \\
Fe XVII  & 323.65 &  0.90 (0.76) & --          & 1.60 (1.66)  \\
\hline
\end{tabular}
\tablefoot{
In brackets the intensity ratios from Shestov et al. (2010) are given.\\
}
\end{table}

\subsection{Hinode/XRT broadband data}\label{sect_43}

We applied the BIM to the same XRT data of Reale et al. (2009), summing them over all pixels within selected regions. The observations were taken on the active region AR10923 observed close to the Sun center on 2006 November 12. The five filters used in our DEM analysis were Al\_poly (B, see Fig. \ref{FigContrFunc}), C\_poly (C), Be\_thin (E), Be\_med (F), and Al\_med (G), with exposure times of 0.26 s, 0.36 s, 1.44 s, 8.19 s, and 16.38 s, respectively. The cadence is five minutes. No major changes in the active region were observed for a full hour between 12:00UT and 13:00UT, and the data from each filter were integrated in this period to increase the signal-to-noise ratio.

\begin{figure*}
   \centering
   \includegraphics[scale=0.8]{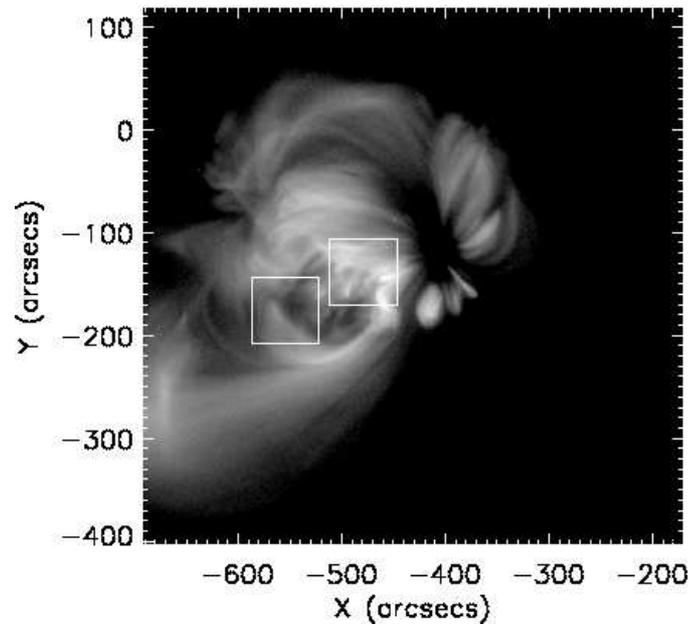}
      \caption{XRT image of the active region AR10923 made with the Al\_poly filter. The boxes (left and right subregions) indicate the regions of different temperature structures.
              }
         \label{FigAR}
   \end{figure*}

Reale et al. (2009) defined two areas (left and right boxes in Fig. \ref{FigAR}) and derived the values of plasma temperature $T_i$ and column emission measure EM($T_i$) corresponding to all numerated by $(i)$ pixels forming these two boxes in different spectral channels (``filters''). These values were obtained by a filter-ratio technique, assuming the isothermal condition in each plasma column along the line of sight. We note that this assumption corresponds to a single-temperature model, described by one singular term $Y_s$ in Eq. (\ref{Ys-term}). For both boxes, the EM($T_j$) histograms for the distribution of emission measure over temperature bins $\Delta T_j$ with a bin size $\Delta\log T=0.1$

\begin{equation}
    \mathrm{EM}(T_j) = \Delta Y(T_j) = \int_{\Delta T_j} y(T)\, dT \, , \label{EM2}
\end{equation}
were built for various pairs of filter ratios, summing up the emission measures EM($T_i$) over all pixels with $T_i\in \Delta T_j$ (see Fig. 5 in Reale et al. 2009)

\begin{equation}
    \mathrm{EM}(T_j)=\sum_i \mathrm{EM}(T_i)=\sum_i Y_s (T_i) \, ,  \qquad            T_i \in \Delta T_j \, .
\end{equation}
All histograms for the left box were found to be similar in shape, being characterized by single-component distributions with
peak temperatures $T_m$ across the temperature range $\log T_m=6.4-6.5$, except for the ``hardest'' F/G filter ratio histogram (F4/F5 in notation of Reale et al. 2009), which also extended to the range $\log T=6.6-7.1$ with a second maximum peaking at $\log T_m = 6.9$ with EM($T_m$) of about 3\% of the maximum EM value of the ``cold'' component. For the right box, all filter ratios were single-component peaked in the range $\log T = 6.5-6.6$.

To explain the absence of a hot component in the F/G ratio histogram for the right box, Reale et al. (2009) suggested that the plasma is probably multi-thermal in both areas (see below dashed histograms in Fig. \ref{Fig_XRTBIM}), but in the right box the hot component is not detected by the filter ratios, because there the main cooler component has a higher peak temperature $T_m$ and completely ``hides'' the minor component.

\begin{figure*}
   \centering
   \includegraphics[width=15cm]{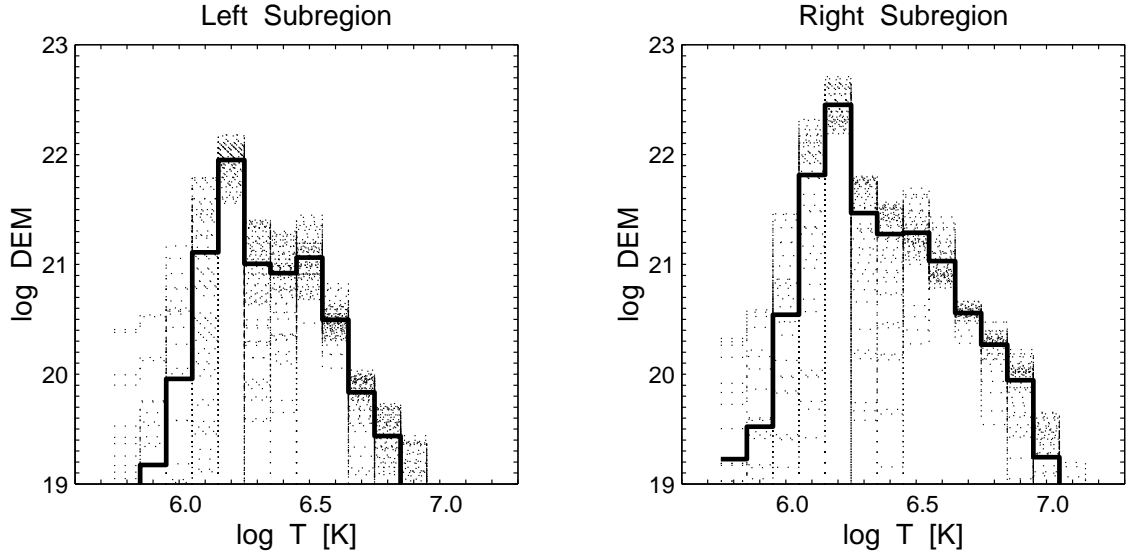}
      \caption{Results of the DEM reconstruction for the AR10923. The heavy solid lines indicate the optimal median DEM solutions; dotted histograms are the solutions of different Monte Carlo statistical realizations. The DEM is in units of [$\mathrm{cm}^{-5}\cdot\mathrm{K}^{-1}$].
              }
         \label{FigDEMXRT}
   \end{figure*}

By applying the DEM($T$) distributions obtained with the BIM algorithm, we have analyzed here the plasma temperature content of the same two boxes identified in Reale et al. (2009) paper. We have used their calibration to perform a compatible analysis and obtained five flux values for each spectral channel (see the list of XRT channels in Table \ref{table:3}), summing up the fluxes over all pixels for both boxes. The resulting DEMs in Figure \ref{FigDEMXRT} are plotted versus the logarithm of temperature using a bin of $\Delta \log T = 0.1$. Here the heavy solid line indicates the optimal median DEMs from the left and right subregions. The dotted histograms indicate the results of the statistical simulation procedure including random variations in the observed fluxes (about 100 different Monte Carlo realizations). The distribution of these noise variations was assumed to be Gaussian with a relative standard deviation of about 3\% for all five channels (see the fourth column in Table \ref{table:3} for the experimental error bars).

\begin{table}
\caption{List of XRT channels used in the DEM analysis of the object AR10923 and ratios of their calculated to observed fluxes.}
\label{table:3}
\centering
\begin{tabular}{c c c c}
\hline\hline
Filter & Calc. / Obs.  & Calc. / Obs. & Exp. error bars,   \\
     &    Left subregion & Right subregion & \% \\
\hline
Al\_poly (B) & $1.02$ (0.94) &  $1.02$ (1.15) &  2.5  \\
C\_poly (C) & $0.98$ (0.88) & $0.97$ (1.07) & 2.6  \\
Be\_thin (E) & $0.98$ (0.74) & $0.98$ (1.26) & 3.0  \\
Be\_med (F) & $1.02$ (0.83) & $1.04$ (1.22) & 3.4  \\
Al\_med (G) & $1.06$ (0.82) & $0.99$ (1.20) & 3.3  \\
\hline
\end{tabular}
\tablefoot{
In brackets the calculated to observed flux ratios for Reale et al. (2009) results are given.\\
}
\end{table}

\begin{figure*}
   \centering
   \includegraphics[width=9cm]{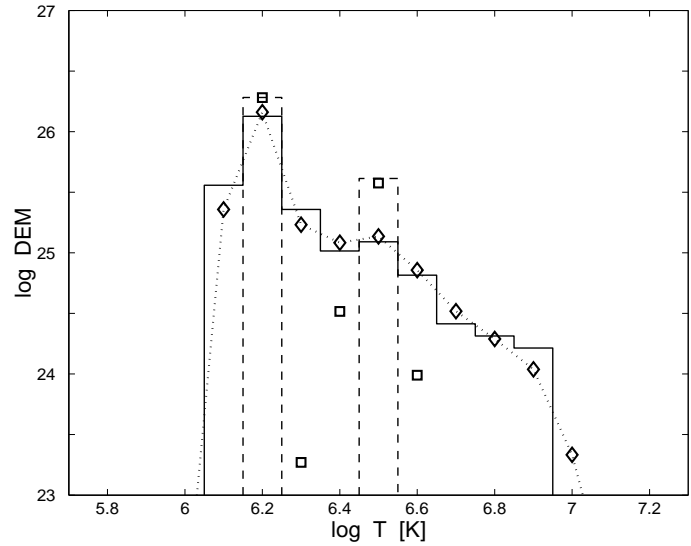}
      \caption{Results of the XRT forward modeling for two input DEM curves: multi-thermal DEM distribution (solid line histogram) and two-temperature DEM with isolated components at temperatures $\log T = 6.2$ and 6.5 (dashed histogram). The diamonds with the dotted line indicate the reconstructed DEM curve for multi-thermal case; the squares are the DEM solution to the two-temperature model. The DEM is in arbitrary units.
              }
         \label{FigDEMXRT_Model}
   \end{figure*}

\begin{figure*}
   \centering
   \includegraphics[width=15cm]{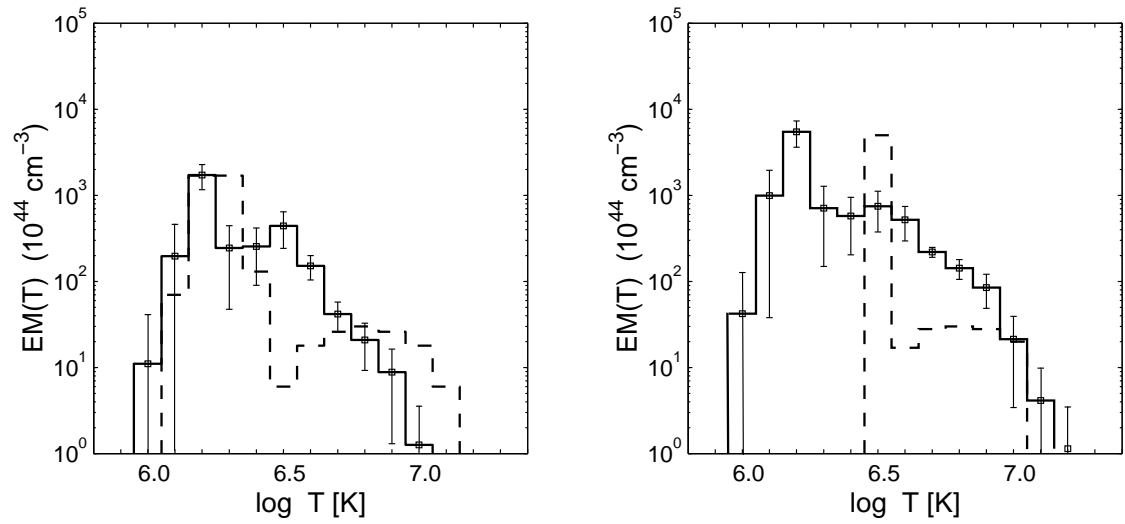}
      \caption{Emission measure distributions vs. temperature, EM($T$), for both left (left plot) and right (right plot) subregions obtained with the BIM algorithm (solid line histograms with $1\sigma$ error bars). Parent EM($T$) distributions obtained from Monte Carlo simulations within the framework of the filter ratio analysis in Reale et al. (2009) are also shown (dashed histograms).
              }
         \label{Fig_XRTBIM}
   \end{figure*}

Figure \ref{FigDEMXRT} shows that the DEMs are similar to each other in shape characterized by a clearly resolved single maximum peaked at $\log T=6.2$, an extended distribution in the range $\log T=6.25-6.55$, and a ``tail'' exponentially decaying at higher temperatures. This reconstructed DEM temperature profile including, apart from a typical quiet Sun cool maximum of $\sim$1.6 MK, an order of magnitude weaker and almost uniform hot component, has not, as far as we are aware, been observed before for ARs. Since, as seen from Figure \ref{FigDEMXRT}, the effective estimator (i.e. characterized by a smaller dispersion) of the reconstructed DEM distributions is in close vicinity to the maximum ($\log T=6.2$) and higher temperatures ($\log T \approx 6.45-6.85$) compared to the intermediate range $\log T \approx 6.2-6.55$, we performed additional forward modeling to assess more thoroughly the quality of the DEM reconstruction. Two input models were used (see Fig. \ref{FigDEMXRT_Model}): a continuously distributed DEM input model (solid line histogram) and a two-component DEM with isolated peaks at $\log T =6.2$ and 6.5 (dashed line histogram). These models were chosen to test ``the resolving power'' of the BIM for an intermediate temperature range. The results of the modeling are shown in Fig. \ref{FigDEMXRT_Model}. As in Section \ref{sect_3}, they indicate the very good agreement between the input (model) DEM and that reconstructed by the BIM, confirming the reliability of the DEM profile in Fig. \ref{FigDEMXRT}.

To compare our results with those by Reale et al. (2009), we calculated the EM($T_j$) histograms using Eq. (\ref{EM2}) and DEM distributions $y(T)$ reconstructed by means of the BIM with corresponding $1\sigma$ error bars estimated according to Eqs. (\ref{mean_value}), (\ref{dispersion}). These histograms, as well as the optimal parent model (OPM) ones (obtained by means of Monte Carlo simulations) adopted by Reale et al. (2009), are shown in Fig. \ref{Fig_XRTBIM}. The calculated-to-observed flux ratios of all channels for histograms obtained by both methods as well as the experimental error bars are presented in Table \ref{table:3}. As follows from Table \ref{table:3}, BIM histograms provide the solutions for both boxes, giving a more reliable, compared to the OPM, quantitative description of observable fluxes (within experimental error bars) in all channels.

As a check of the consistency with the results of Reale et al. (2009),
we calculated the synthetic fluxes of the F and G channels in both boxes using the BIM histograms and inferred the temperatures from the filter ratio F/G. We found $\log T \approx 6.9$ for the left box and $\log T \approx 6.5$ for the right one, with confidence ranges, respectively, of $\log T \approx 6.55-7.1$
and about $\log T \approx 6.45-6.75$ in agreement with the corresponding histograms obtained by the filter ratio method in Reale et al. (2009) (see Fig. 5 of that paper).

We also used the BIM solutions to estimate the quantity of hot plasma EM at $T >$ 4 MK relative to that within the temperature range $\approx$ 2--4 MK corresponding to the AR plasma. This analysis shows that the EM quantity in the range $\approx$ 4--6 MK is $\sim$15\% for the left box and $\sim$30\% for the right one of the EM quantities at $\approx$ 2--4 MK. The upper limit to the EM quantity at the temperature range $\approx$ 9--11 MK is $\approx$ 0.5\% of that for $\approx$ 2--4 MK plasma (for $2\sigma$ one-side confidence interval, i.e. $\approx$ 97.7\% of confidence probability).

\section{Discussion and conclusions}\label{sect_5}

We have described a probabilistic approach to the multi-temperature analysis of line spectra and broadband channel imaging data. The principal feature of this approach is that it uses the language of the probability theory and mathematical statistics, formulating the spectral inverse problem in terms of distribution functions, hypotheses, confidence level etc. It is also essential that the iterative procedure of the BIM is not ``ill-posed'' (in particular, that any fine temperature grid needed, for example, to derive a smooth enough DEM curve be used without creating instabilities in the solution) and, in contrast to many other methods, has been derived in a regular way on the basis of Bayes' theorem.

We applied the BIM algorithm to the DEM analysis of line spectra and broadband channel imaging data. In our studies, we demonstrated that the BIM is an effective means of reconstructing the DEM temperature profiles. We also tested the BIM in the case where we had limited information from observable data.

To demonstrate abilities and robustness of the BIM algorithm, we carried out many tests such as forward modeling and the statistical simulation of input ``experimental'' data. It was established that the BIM provides very good results when analyzing line spectral data, even for a limited number of lines.

We also showed the capabilities of the BIM in the case of data from the broadband multi-filter Hinode/XRT instrument in the temperature range $6.0\lesssim \log T\lesssim 7.0 $. We tested the BIM using both all nine filters of the XRT instrument and the restricted case of the five ones. We showed that in general all temperature plasma components are well identified, both in the case of narrow isothermal and wider multi-thermal distributions. However, for a number of cases the data taken with five filters only do not allow us to reproduce quantitatively the original DEMs at the same confidence level as those with nine filters. In particular, we identified a few exceptions where some small artifacts are usually introduced in the form of small cooler peaks. Throughout our tests, this happened systematically at given temperatures and it was more evident when using the limited number of filters. To understand whether this behavior is a particularity of either the BIM or the XRT instrument filters, we repeated the same tests reported in Sect. \ref{sect_3} using the nine SPIRIT spectral lines. In this case, the artifacts  disappeared. We conclude that some limitations of the BIM results are caused by the relative shape of the XRT response functions of the different filters and their large temperature widths.

We have then applied the BIM algorithm to real observations, which were analyzed previously with other DEM diagnostic techniques. In this case, we had access to elements of comparison. We analyzed the following AR data: SUMER/SOHO (Landi \& Feldman 2008) and SPIRIT/CORONAS-F (Shestov et al. 2010) line spectra data, and XRT/Hinode broadband imaging data. The comparison of the BIM calculations with SUMER/SOHO DEM analysis by Landi \& Feldman showed a very good agreement and confirmed these authors main results and conclusions concerning the coronal thermal structure of the AR considered. For the SPIRIT spectral data, we revised and validated the results of Shestov et al. (2010) including the inference of ARs with different DEM distributions and of hot coronal plasma in the considered ARs at temperatures $\log T \sim 6.8-7.2$.

We also carried out the DEM analysis for XRT broadband imaging data. The BIM approach was applied to the same XRT data analyzed by means of the filter ratio technique by Reale et al. (2009) for a non-flaring AR observed on 2006 November 12. Using data from five filters of the soft X-ray telescope onboard the Hinode satellite, we obtained the DEM curves in two separate areas of the AR. We have shown that they have a similar shape, both with a sharp peak at $\log T \approx 6.2$ (typically the quiet Sun structure) and with a significant broader component around $\log T \approx 6.5-6.6$ (active region plasma). The BIM calculations also detected hot plasma in the two considered AR subregions. The amount of EM in the temperature range $\approx$ 4--6 MK in comparison to that at $\approx$ 2--4 MK is about $\sim$15\%  and 30\% for, respectively, the left and right subregions of the AR under study.

In comparison with the optimal parent model (OPM) histograms proposed by Reale et al. (2009), the histograms restored by the BIM along with our simulations and data analysis contained distributions of the same type for both boxes with a prominent peak at $\log T=6.2$, a significant bump around $\log T \approx 6.5$, and a decreasing tail at higher temperatures. This noticeable difference between the histograms indicates that more studies of possible solutions for the five channel XRT data are needed.

To our knowledge, very few studies have been made where the DEM was derived in active regions using soft X-ray data. One of these (Schmelz et al. 2009a, 2009b) used the standard xrt\_dem\_iterative2.pro procedure in Solarsoft.
It would be interesting to compare the BIM results with those obtained using that procedure. Unfortunately, the number of XRT filters used for our observations are insufficient to make the Solarsoft procedure work correctly. We plan to perform this test in the near future using different XRT data.

\begin{acknowledgements}
      FG, SP and JFH acknowledge the support from the Belgian Federal Science Policy Office through the international cooperation programmes  and the ESA-PRODEX programme. This work was also supported by the European Commission's Seventh Framework Programme (FP7/2007-2013) under the grant agreement 218816 (SOTERIA project, www.soteria-space.eu), the grant of the Russian Foundation of Basic Research 08-02-01301-a and ``Precision optical spectroscopy of interatomic and intermolecular transitions'' -- Program of the Department of Physical Sciences of the Russian Academy of Sciences. We are also grateful to Enrico Landi for providing us with the DEM curve from Landi \& Feldman (2008) paper. Hinode is a Japanese mission developed and launched by ISAS/JAXA, with NAOJ as domestic partner and NASA and STFC (UK) as international partners. It is operated by these agencies in cooperation with ESA and NSC (Norway). CHIANTI is a collaborative project involving researchers at NRL (USA) RAL (UK), and the Universities of: Cambridge (UK), George Mason (USA), and Florence (Italy). FR acknowledges support from Italian Ministero dell'Universit\`a e Ricerca and Agenzia Spaziale Italiana (ASI), contract I/023/09/0. The authors would like also to acknowledge the anonymous referee for a deep interest to the problem which helped to significantly improve the clarity of the manuscript.
\end{acknowledgements}

%

%

%

%

%

%

%

%

%

%

\end{document}